\documentstyle[epsfig,rotating]{jaa}

\DeclareMathAlphabet{\mathsc}{OT1}{cmr}{m}{sc}
\def\testbx{bx}%
\DeclareRobustCommand{\ion}[2]{%
\relax\ifmmode
\ifx\testbx\f@series
{\mathbf{#1\,\mathsc{#2}}}\else
{\mathrm{#1\,\mathsc{#2}}}\fi
\else\textup{#1\,{\mdseries\textsc{#2}}}%
\fi}
\begin{document}
\title[Core-collapse Supernovae and Gamma-ray Bursts in TMT Era]
{Core-collapse Supernovae and Gamma-ray Bursts in TMT Era}
\author[S.~B.~Pandey] 
{S.~B.~Pandey$^{1}$\thanks{e-mail: shashi@aries.res.in} \\
(1) Aryabhatta Research Institute of observational-sciences, Manora peak, 
\newline
Nainital 263 129, India\\
}

\pubyear{xxxx}
\volume{xx}
\date{Received xxx; accepted xxx}
\maketitle
\label{firstpage}
\begin{abstract}

Study of energetic cosmic explosions as a part of time domain astronomy 
is one of the 
key areas that could be pursued with upcoming Giant segmented optical-IR
telescopes with a very large photon collecting area applying cutting edge 
technology. Existing 8-10m class telescopes have been helpful to improve our 
knowledge about Core-Collapse Supernovae, Gamma-ray Bursts 
and nature of their progenitors and explosion mechanisms. However, many 
aspects about these energetic cosmic explosions are still not well-understood 
and require much bigger telescopes and back-end instruments with high 
precision to address the evolution of massive stars and high-redshift Universe
in more detail. In this presentation, possible thrust research areas towards 
Core-Collapse Supernovae and Gamma-ray Bursts with the Thirty Meter Telescope 
and back-end instruments are presented.  

\end{abstract} 


\section{Introduction}

The astronomical community with the aid of ground/space based telescopes 
has made tremendous progress over the last hundred years or so, in 
understanding many aspects of our observable universe. The 
findings include: discovery of exo-planetary systems, evidence for an 
accelerating universe, detailed identification and monitoring of the orbits of 
the asteroids and comets that may pose great dangers to the inhabitants of the 
Earth, and many more. Astronomers across the world feel a great need for new 
generation very large telescopes to probe the Universe much deeper, to unravel 
its formation and evolution, and to discover the existence of life elsewhere 
in the Universe. With the current combination of 8-10m class ground-based 
telescopes and the Hubble Space Telescope, study about Core-Collapse 
Supernovae (CCSNe) and Gamma-ray Bursts (GRBs) have been able to provide 
a great deal of information about the fate of evolution of massive stars 
($> 8-10 M\odot$) and the underlying physical mechanisms (Woosley \& Bloom 
2006; Langer 2012 and references therein). 
The unprecedented light gathering power and spatial resolution of the 
Thirty Meter Telescope\footnote{www.tmt.org/documents} (TMT) and the back-end 
instruments\footnote{http://www.tmt.org/observatory/instruments} will allow us 
to study distant lighthouses like CCSNe 
and GRBs, could be studied in more detail probing the high-redshift
Universe, dark energy contents and epoch of re-ionization.

\subsection{The TMT Project}

The TMT project is one of the most innovative among 
the proposed mega-science projects in the next decade. The project is aimed 
to build the world's most advanced and capable ground-based optical, 
near-infrared, and mid-infrared observatory. It will integrate the latest 
innovations in precision control, segmented mirror design, and adaptive optics 
(AO). At the heart of the telescope is the segmented mirror, made up of 492 
individual segments of 1.45m. Precisely aligned, these segments will work as 
a single reflective surface of 30m diameter. TMT will be 
located just below 
the summit of Mauna Kea on Hawaii Island, at an elevation of 4050 meters. The 
performance of a ground based telescope is adversely affected by the Earth's 
atmospheric seeing. The fundamental goal of any AO system is to improve the 
telescope performance from seeing limited, meaning the image quality is 
limited by the atmosphere above the telescope, toward diffraction limited, 
meaning images as sharp as those that could be obtained with the same diameter 
telescope located in space. Even in seeing limited mode, TMT offers an order 
of magnitude improvement over existing observatories, mostly due to its light 
gathering capacity. The AO capability enables TMT to resolve objects by a 
factor of 3 better than the current 10-m class telescopes and 12 times better 
than the Hubble Space Telescope. As first generation instruments for TMT, The 
Wide Field Optical Spectrometer (WFOS) will provide near-ultraviolet and 
optical (0.3 to 1.0 $\mu$m) imaging and spectroscopy over a more 
than 40 square arc-minute field-of-view. The concept for the TMT Wide-Field 
Optical Spectrometer is the Multi-Object Broadband Imaging Echellette (MOBIE) 
spectrometer. Narrow Field Infrared Adaptive Optics System (NFIRAOS) is the 
TMT's adaptive optics system for infrared instruments (0.8 - 2.5 $\mu$m). 
Two of these science instruments will be delivered for use with the Multi 
Conjugate Adaptive Optics-Laser Guide star (MCAO-LGS) system at first light: 
IRIS-TMT, a near-infrared instrument with parallel imaging and 
integral-field-spectroscopy support; and IRMS, an imaging, multi-slit 
near-infrared instrument (Crompton, Simard \& Silva 2008, TMT\_documents1,2,3). 
The suit of above first generation instruments with TMT along with synergy 
with the {\it E-ELT}\footnote{http://www.eso.org/public/teles-instr/e-elt.html} and {\it JWST}\footnote{http://www.jwst.nasa.gov/} would also be 
very relevant studying CCSNe and GRBs in great details. 

\subsection{Indian Historical Prospects} 

India has made several notable contributions towards optical-NIR astronomy 
during the latter half of the last century and had put in great efforts to set 
up world class observing facilities, which culminated in the indigenous 
building of the 2.3m Vainu Bappu Telescope (VBT) in 1987. The most recent 
astronomy facilities which have been set up in the country are, IIA's 2.0m 
Himalayan Chandra Telescope (2003) at Hanle, Ladakh and the 2.0m IUCAA 
Girawali telescope (2006) at Girawali, near Pune. There are two upcoming 4.0m 
class optical-NIR astronomical facilities led by ARIES, Nainital: the 3.6m 
optical-NIR facility (Devasthal Optical Telescope) which is expected to be 
commissioned in 2013, and the proposed 4.0m International Liquid Mirror 
Telescope, both being set-up at Devasthal (Sagar et al. 2012). 
India will soon launch a dedicated astronomy satellite, {\it ASTROSAT}, with 
multi-wavelength capabilities in the UV and X-ray wavelengths. Indian 
astronomers have also contributed towards studying 
stellar explosions and early universe apart from other areas of observational 
astronomy. Using the observational capabilities in the country during last two 
decades, Indian astronomers have been able to study many SNe (e.g. Ashok B N 
et al. 1987; Pandey et al. 2003; Sahu et al. 2006; Roy et al. 2008; 
Brajesh et al. 2013) and afterglows of GRBs (e.g. Sagar et al. 1999; 
Bhattacharya 2003; Pandey et al. 2003, 2004; Resmi et al. 2013) in detail.

\subsection{India-TMT}

India's participation in international projects was envisaged in the Astronomy 
and Astrophysics 'Decadal Vision Document 2004' sponsored by the Indian Academy
of Sciences and the Astronomical Society of India. In this context, 
international consortia for mega telescope projects approached astronomy
institutes in the country for India's participation. The Indian astronomy 
community after due diligence and thorough deliberations, arrived at the 
conclusion that TMT presents the best options for India and participation 
in the project at a 10\% level would be optimal. 

The Department of Science and Technology (DST) reviewed the proposal submitted 
by Indian astronomers and approved the observer status for India in the TMT 
project in June 2010. Since then, India has been participating in all the 
policy decisions and development activities of the project.
The Aryabhatta Research Institute for Observational Sciences (ARIES), Nainital; 
the Indian Institute of Astrophysics (IIA), Bangalore; and the Inter-University
 Center for Astronomy and Astrophysics (IUCAA), Pune; are the three main 
institutes spearheading the efforts. Options for other Indian Institutes and Universities
are open to join the ongoing efforts by India-TMT. The activities of India-TMT\footnote
{tmt.iiap.res.in} will be 
coordinated by the India TMT Coordination Center (ITCC), jointly funded by 
the Department of Science and Technology and Department of Atomic Energy, 
Government of India.

\section{Core Collapse Supernovae}

It is commonly recognized that CCSNe represent the final stages of the life 
of massive stars ($M >$ 8\,--10 M$_{\odot}$) (Heger, Fryer \& Woosley 
2003; Anderson \& James 2009). Generally, the 
fate of massive stars is governed by its mass, metallicity, rotation and 
magnetic field (Fryer 1999; Woosley \& Janka 2005). Massive stars 
show a wide variety in these fundamental parameters, 
causing diverse observational properties among various types of CCSNe.
The presence of dominant H lines in the spectra of Type II SNe strongly 
suggests that their progenitors belong to massive stars which are still 
surrounded by significantly thick hydrogen envelope before the explosion 
(Filippenko 1997). For a Salpeter IMF with an upper cut-off at 
100 M$_{\odot}$, half of all Type II SNe are produced by stars with masses 
between 8 and 13 M$_{\odot}$. This means that more than half of the stars 
producing SNe are poor sources of ionizing photons and of UV continuum 
photons, and that the bulk of the UV radiation, both in the Balmer and in 
the Lyman continuum is produced by much more massive stars. So, both the 
H-alpha flux and the UV flux are measurements of the very upper part of the 
IMF (about $> 40 M_{\odot}$), representing stars with masses larger than 8 
M$_{\odot}$ and are least understood. The analysis of archival images of 
{\it HST} have been able to detect some of the type II SNe  progenitors as 
red and yellow super-giants (Yoon et al. 2012, Van dyk et al. 2012). 
On the contrary, H and He deficient features are commonly observed in the  
spectra of Type Ib/c SNe and are supposed to have luminous Wolf-Rayet stars
as the possible progenitors. However, search for Type Ib/c SNe  progenitors in 
the archival deep images  has thus far been unsuccessful.  
The rate of core-collapse SNe (II, Ib/c) is a direct measurement of the 
death of stars more massive than 8 M$_{\odot}$, although it is still a 
matter of debate whether stars with mass above 40 M$_{\odot}$ produce 
a 'normal' SNe-II and Ib/c, or rather collapse forming a black hole 
with no explosion, i.e. a ``collapsar'' (e.g. Heger \& Woosley 2002). 
These explosions and their possible progenitors are rather poorly understood 
research problems in astrophysics and a subject of great scientific 
interests (Taubenberger et al. 2009; Modjaz et al. 2011; Crowther 2013).

%%%%%%%%%%%%%%%%%%%%%%%%%%%%%%%%%%%%%%%%%%%%%%%%%%%%%%%
\begin{figure}
\centering
\includegraphics[width=9cm, angle=0]{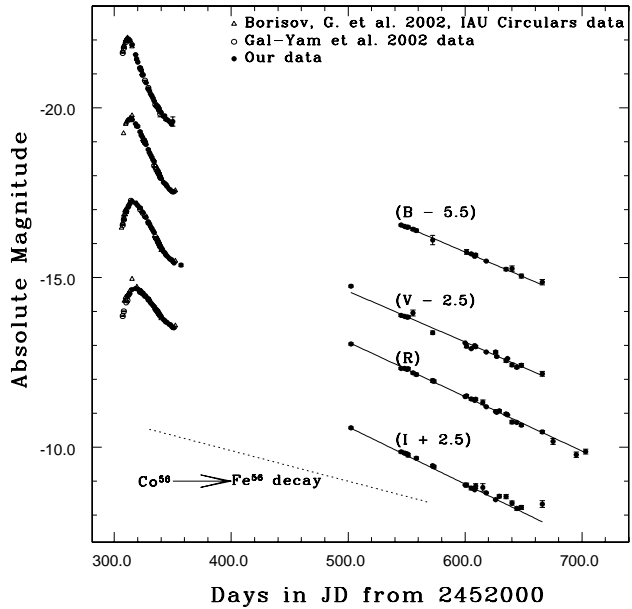}
\caption{$BVRI$ light-curves of broad-lined Type Ic SN 2002ap discovered in M74 at 
z $sim$ 0.002, showing the typical decay nature of H-stripped CCSNe (Pandey 2006).
SN 2002ap is more energetic than typical SNe of Type Ib/c and exhibit close
resemblance with SN 1998bw profile at later epochs.}    
%\label{fig:Sample}
\end{figure}
%%%%%%%%%%%%%%%%%%%%%%%%%%%%%%%%%%%%%%%%%%%%%%%%%%%%%%%

At low (z $<$ 0.1), intermediate (0.1 $<$ z $<$ 1.0) and high (z $>$ 1.0) 
redshifts many 
SN search programs  such as {\it CRST, PTF, KAIT, CfA, ESSANCE, SDSS, HST-GOODS} 
are going on whereas many more are proposed to some in coming years e.g.  
{\it SKYMAPPER, PANSTARRS, LSST, WFIRST, JWST}. So far, the fraction of total 
number of discovered CCSNe ($\sim$76\%) is found to be much more in comparison to 
the number of type Ia SNe ($\sim$24\%). Out of total number of discovered CCSNe, 
Type-II events are much more than Type Ib/c SNe (Lennarz, Altmann \&  
Wiebusch 2012). Statistically, the number of Type Ib/c SNe in the local 
universe have been very less (only 4\% of the total population) in 
comparison to number density of Type-II SNe and their possible progenitors 
are WR stars. No conclusive results have been found for the brightness 
distribution of CCNSe with their redshifts (Habergham, James \& Anderson 2012).

The TMT along with the back-end instruments is optimized both in OA and 
seeing-limited modes for the rapid response i.e. 10 minutes, extremely 
important for studies of time critical observations like SNe and GRBs. 
For example, IRIS-TMT spectra of a point source in one hour exposure time with 
S/N $\sim$ 10 will be able to look at J $\sim$ 24.1 mag, H $\sim$ 23.7 mag 
and K $\sim$ 22.9 mag. Whereas seeing limited imaging of S/N $\sim$ 100 
will be bale to see sources upto J $\sim$ 27.3 mag, H $\sim$ 26.2 mag and 
K $\sim$ 25.5 mag in a similar exposure time.     

\subsection{CCSNe of Type II by TMT}

SNe IIn (~ 3\% of type-II populations) are the brightest SN type in the far-UV 
and have long lived emission lines, makes them potential candidates to study 
in great detail at high redshifts. The average far-UV flux of Type IIn SNe at 
z $\sim$ 2 corresponds to m\_v $\sim$ 28 mag, enabled photometric detections in deep 
surveys. Moreover, the bright long-lived spectral features seen in case of 
Type-IIn SNe, remain above the threshold of 8-10m class telescope for years 
after the explosion, enabling WFOS-TMT kind of instrument to obtain rest-frame 
far-UV emission-line detections, spectral properties, and inferred kinematics 
of z $>$ 2 Type IIn SNe and host galaxies. The rest-frame far-UV light would be
red-shifted to optical wavelengths for 2 $<$ z $<$ 6 SNe and enables photometric 
detection in existing, and future, deep (m\_v $\sim$ 27-28 mag) multi-epoch, 
wide-field optical surveys using 4-8m-class facilities. The ejecta of SNe 
IIn interact with cold circumstellar material expelled during previous 
evolutionary episodes and create extremely bright and long-lived emission 
lines. Current 8-10m class telescopes do not have spectral sensitivity to 
detect Ly-alpha emission from z $>$ 3 SNe of Type II. The strength and duration 
of the prominent emission lines permit spectroscopic detection of 2 $<$ z $<$ 6 
using IRMS and WFOS-TMT for $\sim$3-15 years after outburst (Cooke et al. 2009; 
Whalen et al. 2013a). Early time spectroscopy of Type IIn SNe during 
photospheric and shock-breakout phases using TMT is very important 
to understand the environments and the nature of the possible progenitors  
(Schawinski et al. 2008, Bufano et al. 2009). Late time spectral 
evolution of Type IIn SNe with TMT will be important to understand the 
geometry of the inner ejecta using evolution of line profiles and chemical 
enrichment of ISM and IGM. 
 
On the other hand, CCSNe Type IIp are quite common in nearby universe and 
their atmosphere are dominated by H lines. Observational constraints on 
the possible progenitors of Type IIp SNe (Smartt 2009) 
using the pre-explosion images further make these locations as promising 
targets for TMT to put limits on the masses and types of the progenitors.   
It is also possible to probe Type IIp SNe at z $>$ 1 to study possible 
shock-breakout phase probing the underlying mechanisms of the possible
progenitors of Type IIp SNe using TMT capabilities (Tominaga et al. 2011). 
Existing 8-10m class telescopes required 2-3 hours of exposure time to get 
a good S/N nebular phase spectra (z $<$ 0.3) with Fe-II lines showing  
correlation with I-band luminosity, not seen in case of other Type II SNe. 
So, with the TMT capabilities, high-z SNe of Type IIp could be studied to
be used as standard candles (Pozananski et al. 2012). 

\subsection{CCSNe of Type Ib/c by TMT}

Hydrogen-stripped Type Ib/c SNe (see Figure 1) are another potential targets for TMT to be
studied in more details. The upcoming SN survey programs like LSST, WFIRST, 
JWST are expected to produce a sizable number of high-z Type Ib/c SNe. Massive 
Wolf-Rayet stars ($>$ 25-30 M${_\odot}$) are supposed to be the potential 
progenitors for these SNe (Crowther 2007), so far not identified in 
per-discovery images taken by {\it HST} and ground-based 8-10m class telescopes 
(Maund, Smartt \& Schweizer 2005; Crockett et al. 2007, 2008). Another 
plausible way to produce Type Ib/c SNe is a lower mass helium-star in a 
binary channel with mass-loss mechanism (Podsiadlowski et al. 1992; 
Nomoto, Iwamoto \& Suzuki 1995). Majority of Type Ib/c SNe are found in 
star forming galaxies. Taking advantage of TMT OA and IFU 
capabilities, stellar population studies (Murphy et al. 2011; Anderson 
et al. 2012, Eldridge et al. 2013) close to the explosion site would tell 
more about the possible progenitors of Type Ib/c SNe.  

Using WFOS-TMT capabilities, late time optical spectroscopy of type Ib/c SNe 
could tell about kinematical, chemical information about the shocked ejecta
and its interaction with the circumstellar material 
(Chevalier \& Fransson 2006). Late time optical spectroscopy could also 
help probing  the mass loss history and evolutionary status of the 
progenitor stars (Leibundgut et al. 1991; Fesen et al. 1999). 
Time series optical spectroscopy of Type Ib/c SNe in nebular 
phases could also tell about the evolution of line widths and ratios of 
various line strengths, distinguishing the late time underlying physical 
mechanisms and ejecta structure. Recent observations of light echoes in 
supernova remnants have opened yet another method to determine the 
progenitors of supernovae. For example, light echoes from SN 1993J have 
been detected at the level of 22.3 mag/arcsec$^2$ square by 
{\it HST} "V" band (F555W) observations. TMT class of telescopes will detect 
light echoes to V ~ 28.5 mag/arcsec$^2$ in galaxies well beyond the Virgo 
Cluster. With this capability one can determine the nature of SN types of 
all recorded explosions whose types (i.e. whether thermonuclear or 
core-collapse) are not known, in the past hundred years or so or of the SN 
remnants in the Local Group of Galaxies by faint object spectroscopy of 
the echoes of the light given out originally by the SN when it exploded 
(Milisavljevic et al. 2012). 

\subsection{Pair instability SNe}

Pair instability supernovae (PISN) are thought to arise from extremely massive 
progenitors, possibly population-III stars above 100 solar mass 
(Rakavy \& Shaviv 1967, Barkat et al. 1967). Massive stars, with an initial 
mass range 140 $<$ M $<$ 260 M${_\odot}$ die in a thermonuclear runaway triggered 
by pair production instability (Kasen et al. 2011; Joggerst et al. 2012). 
Although PISN are rarely observed in the present, it is thought that they 
were very numerous in the distant past, among primordial, super-massive, 
low-metallicity rotating stars (Chatzopoulos \& Wheeler 2012), releasing 
huge energy of the order 10$^53$ ergs and may synthesize considerable 
amount of 56Ni (upto 50 M${_\odot}$).
The PISN are characterized by peak magnitudes that are brighter than of 
Type II SNe and comparable, or brighter than type Ia, fall in category of 
super-luminous SNe of Type I and II observed in nearby universe 
(Quimby et al. 2013). Hydrogen lines are present in these SNe, 
arising from the outer envelope and have a long decay time ($\sim$1 year) due 
to large initial radii and large mass of material involved in the explosion. 
Very massive stars that can retain most of their mass against mass loss 
are those that with metallicity close to zero, and are hence expected 
at high redshifts. At redshift z $\sim$ 2, the PISN/SN ratio is expected to be
4-10 times higher than the observed local rate, for sub-solar mettalicity stars 
(Langer \& Norman 2006, Langer et al. 2007). Study of PISN using IRIS-TMT will 
also help in understanding the history of chemical enrichment, the nature 
of metal free stars at z $\sim$ 6 and the evolution of gaseous matter in the 
Universe (Scannapieco et al. 2005, Whalen et al. 2013b,c). 

\subsection{Environment and Progenitors of CCSNe}

Host galaxies of CCSNe provide very useful information about the environments 
of these explosions, indirect clues about nature of their progenitors and 
evolution of high mass stars. With the help of ground-based 8-10m class 
telescopes and {\it HST} imaging capabilities with a differential alignments of 
10-130 mas, progenitors of some of the Type-II SNe have been identified, 
though limited to 10-120 Mpc only (Smartt et al. 2009;  Modjaz et al. 2011). 
There are evidences, though not conclusive, for a correlation of masses of 
the progenitors with the types of CCSNe (Crowther 2012). However, estimates of 
host mettalicities does not show any such conclusive difference for various 
types 
of CCSNe (Boissier \& Prantzos 2009; Modjaz et al. 2011; Kelly and Krishner 
2011). Locations of Type Ib/c SNe seems to be more centrally concentrated 
than Type II SNe in their host galaxies (Anderson \& James 2009; Arcavi et 
al. 2010).    

With the help of AO capabilities of TMT and first generation instruments like 
WFOS and IRIS-IFU, mettalicies and star formation rate of the immediate 
environments of the host galaxies could be studied on the scales of kpc 
(Modjaz et al. 2008; Sahu et al. 2009; Anderson et al. 2010; 
Modjaz et al. 2011; Neill et al. 2011), providing more 
conclusive results about ages, mettalicity of the stellar population. 
TMT-IRIS and WFOS could be used to obtain rest-frame far-UV emission-line 
detections, spectral properties, and inferred kinematics of a large number 
of  z $>$ 2 galaxies hosting CCSNe of Type II and Ib/c, useful to understand 
the density, evolution, and dynamics of such events and the process of 
galaxy formation. There are evidences that some of the SNe Type IIn are 
PISN events (Langer 2012; Whalen et al. 2013a), providing a clue to understand 
the evolution of 
pop-III stars using follow-up observations of a larger sample of z $>$ 2 
Type IIn SNe by LSST and other upcoming surveys.   

\section{Gamma-ray Bursts}

GRBs are short lived ($10^{-3}$ to $ 10^{3}$ seconds) 
extremely bright (Isotropic equivalent $\gamma-$ray energy 
$\sim 10^{50} - 10^{55}$ erg) cosmological (redshift z $\sim$ 0.01 to 8.3) 
$\gamma-$ray sources, emitting photons of energy $\sim$ 10 keV--90 GeV. 
They appear at random locations in the sky and have non-thermal spectrum
(Piran 1999; Gehrels, Ramirez-Ruiz \& Fox 2009). Long-duration
GRBs are now believed to be relativistic analogues of CCSNe explosions 
and are among the most energetic stellar-scale events known so far 
(Woosley \& Bloom 2006). The gamma-rays are not effected by dust absorption 
and optical extinction, hence could be seen at very high redshifts 
(Lamb \& Reichart 2000). Since last more than a decade, we have been able to 
understand GRBs and their properties using all possible ground and space-based 
observing facilities, however, many more questions about these enigmatic 
explosions are yet to be answered (Zhang 2007). GRBs are the only astrophysical sources 
known to be found in low-mettalicity environments and tools to probe the 
progenitors at high redshifts. Capabilities of TMT along with the first 
generation instruments specially the AO system with IRMS will be very helpful 
exploring high-redshift universe using GRBs.

\subsection{Afterglow light curves at late phases}

Followed by the prompt emission, ultra-relativistically ejected material 
interact with 
the surrounding medium through shocks and may produce afterglows, visible in 
all bands from $X-$ray to radio wavelengths. Afterglows being longer-lasting 
than GRB prompt emission, provide a multi-band platform to study these 
energetic cosmic explosions in detail. The afterglow light curves generally 
show a power-law behavior. In general the flux $f_\nu$ from the afterglow 
follows a power law decay with time, a combination of several characteristic 
properties of the ejecta, represented as 
$f(\nu,t) \propto \nu^{_\beta}t^{-\alpha}$ where $t$ is the time since the 
burst and $\alpha$ is the temporal flux decay index, $\nu$ is the frequency 
of observations and $\beta$ is the spectral index (Sari, Piran \& Narayan 1998, 
Wijers \& Galama 1999; Pandey et al. 2004). In case of most of the afterglows, 
generally the optical-IR emission is very faint (V $\sim$ 23 mag, 1-2 
day after the burst) and only 8-10m class telescopes could monitor the 
emission to observe the characteristic jet break time and other possible 
light curve features at late epochs (Sari, Piran \& Halpern 1999;
; Castro-Tirado et al. 1999; M\'esz\'aros 2002). Scaling similar bursts at z $\sim$ 1 to 
z $\sim$ 6 - 10, the expected jet-break time and other light curve features 
(see Figure 2) would shift at later epochs and the 
afterglow emission would also shift towards IR bands due to Ly-alpha 
absorption, making the observations difficult even for existing 8-10m class 
telescopes. TMT imaging capabilities will be able to detect a good number 
of afterglows up to z $\sim$ 6 - 10 with the help of upcoming 
next generation GRB missions like {\it EXIST}. 

One third of GRBs, terms as "dark GRBs", do not show any optical-IR emissions 
to a very deep limits at very early epochs even observed with 8-10m class 
telescopes. The afterglow observations of these bursts at X-ray and 
mm-wavelengths indicate that these bursts could either be extinguished by 
heavy dust of the host galaxies or exploded in a very low-density 
environments (Reichart \& Price 2002). It is also proposed that 
these bursts 
lie at very high redshifts (Lamb \& Reichart 2000) or these bursts are 
intrinsically very faint (Fynbo et al. 2001; Berger et al. 2002: 
Malendri et al. 2012: van der Horst et al. 2009: Jakobasson et al. 
2004). Deep imaging capabilities of WFOS/IRIS-TMT with AO will be very 
useful to answer these questions about "dark GRBs".

\subsection{Time-resolved spectroscopy of afterglows}

Spectroscopy of afterglows of GRBs have many more useful implications than 
just measuring the distances. Early time afterglow spectra provide information 
about the absorption features due to fine structure transitions and resonance 
enabling accurate measurements about the dust content, chemical composition 
and complex gas kinematics of host and other intervening galaxies in the line 
of sight (Savaglio 2006, Prochaska et al. 2008). Present 8-10m class telescopes need lot of time to get spectra of GRB afterglows, for example, Subaru-FOCAS 
required $\sim$ 4 hours for GRB 050904 (z=6.295), VLT-FORS2/ISAAC required 
$\sim$ 1.0 hour 
for GRB 090423 (z=8.3) to get spectra just for the redshift determination 
(Totani et al. 2006; Tanvir et al. 2009). Gravitational collapse of 
rapidly evolving massive stars are supposed to produce long-duration GRBs, 
useful to probe young star forming regions to quite far away distances 
(Tanvir et al. 2009). In fact, spectroscopy of 
optical-IR afterglows 
are proxies to probe the interstellar medium and intergalactic gas at 
cosmological distances in detail (Vreeswijk et al. 2007, 2013; D'Elia et al. 
2009; Ledoux et al. 2009). Time resolved spectroscopy of optical-IR 
afterglows could also be used to understand the collisional excitation of 
fine structure lines, dust destruction effects due to UV/X-ray afterglows, 
measuring ISM densities and abundances (Vreeswijk et al. 2007). Taking 
advantage of  rapid response modes along with the suite of back-end 
spectroscopic facilities (wavelength range = 0.31-0.62 $\mu$m \& 2-2.4 $\mu$m; 
R=1000-50000 and 0.05 mas astrometry), TMT would be able to address 
many of these aspects 
in detail not possible to understand with help of current 8-10m class 
telescopes. High resolution spectroscopy of GRBs with TMT-HRMS 
(at 0.34-1.0 $\mu$m, 
SN/100, m\_{AB} $\sim$ 20) could be able to provide information about components 
effecting ISM upto kpc scales. Currently, the distances of absorbing gas 
from GRBs and study of fine structure lines have been done just in handful of
cases only using VLT-MISTICI and X-SHOOTER instruments (D\_Elia et al. 2007, 2009, 2011 
; Ledoux et al. 2009; Piranomonte et al. 2008; Vreeswijk et al. 2007, 2013).

\subsection{Observational cosmology and GRBs} 

GRBs at high-z could be used as proxies to high-z quasars to 
investigate the epoch of re-ionization. As the end products of massive 
stars, GRBs briefly outshine any other source in the 
Universe and can be easily observed at cosmological distances. The recent 
discovery of a GRB at a redshift of z $\sim$ 8.3 (Tanvir et al. 2009, 
Salvaterra et al. 2009), around similar redshifts of the most distant 
spectroscopically confirmed galaxies and quasars, establishes that 
massive stars 
were being produced and dying at these epochs. This recent detection of the 
distant GRBs using 4-10 m class telescopes clearly indicates that it is 
possible to detect and study many GRBs at z $>$ 10 using the TMT and 
back-end instruments. GRBs being associated with 
individual stars, serve as signposts of star formation at the early epochs and
can also provide a measure of the neutral fraction of IGM at the location 
of the burst. The observations of several GRBs at high-z would provide multiple 
lines of sight through the IGM and thus allow us to trace the process of 
re-ionization from its early stages. 

%%%%%%%%%%%%%%%%%%%%%%%%%%%%%%%%%%%%%%%%%%%%%%%%%%%%%%%
\begin{figure}
\centering
\includegraphics[width=13cm, angle=0]{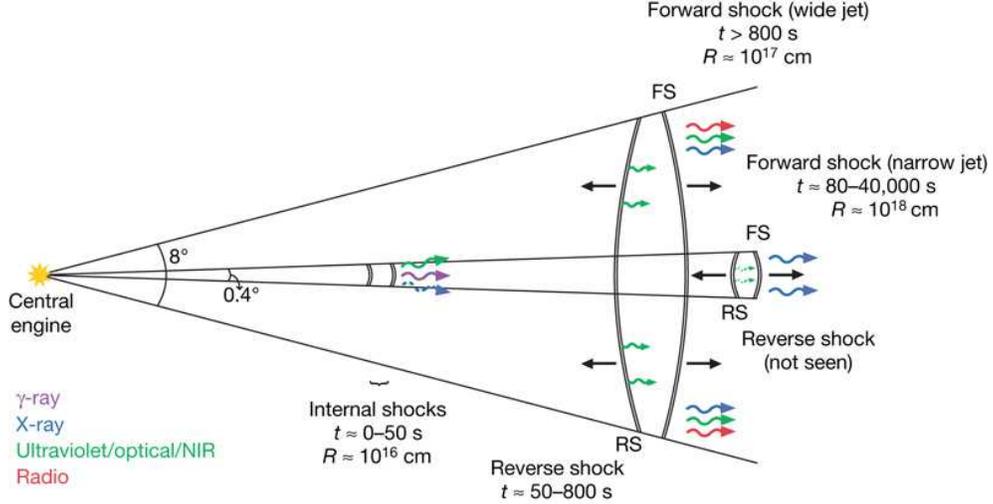}
\caption{Schematic of the proposed two-component jet model of the afterglow of the 
``Naked-eye'' GRB 080319B. The inner narrow-jet gives rise to the high energy emission 
whereas wider-jet is supposed to produce forward shock emission (Racusin et al. 2008).
GRB 080319B (z $\sim$ 0.94) was very bright at optical frequencies and could have 
been observed even at redshift z $\sim$ 16 with the existing IR facilities hours after 
the burst.}
%\label{fig:Sample}
\end{figure}
%%%%%%%%%%%%%%%%%%%%%%%%%%%%%%%%%%%%%%%%%%%%%%%%%%%%%%%

Above properties suggest that GRBs could be good candidates to probe 
the cosmological models of the Universe with a much longer arm than Type Ia 
SNe (Soderberg et al. 2006, 2010). However, the huge dispersion 
(four orders of magnitude) of the isotropic GRB energy E$_{iso}$ makes them 
everything but standard candles barring the clustering of energy around 
E$_{jet} \sim 10^{51}$ erg corrected for their jet opening angle (Frail et al. 2001) 
with a considerable scatter. The recent discovery of a very tight correlation 
between the collimation corrected energy E$_{jet}$ and the GRB spectral peak energy 
E$_{peak}$ are used as standard candles to constrain the cosmological parameters 
(Ghirlanda et al. 2004). The prospects for the use of GRBs as standard 
candles clearly depend on the increase of the number of detected GRBs which 
satisfy the E$_{peak}$-E$_{jet}$ correlation for a much longer redshift span. This will 
offer the unprecedented opportunity to investigate the nature of dark energy 
beyond what can be reached by study of Type Ia SNe. Cosmology with GRBs 
through the E$_{peak}$-E$_{jet}$` 
correlation requires a set of observables, which are derived both from the GRB 
high energy emission (i.e. the prompt gamma-ray emission phase) and from the 
afterglow observations in the optical and IR band. In particular, the afterglow 
spectroscopic observation should provide the GRB redshift, while the long 
term (days to weeks) photometric monitoring of the afterglow emission allows 
measurement of the jet-break time. The latter allows us to estimate the GRB 
opening angle $\theta_{jet}$ and, therefore, to derive the collimation 
corrected energy E$_{jet}$. TMT imaging capabilities at IR bands would be 
quite helpful constraining jet-break time in conjunction with facilities at 
other wavelength for very high redshift GRBs.`

\subsection{Supernova connection of GRBs}

%%%%%%%%%%%%%%%%%%%%%%%%%%%%%%%%%%%%%%%%%%%%%%%%%%%%%%%
\begin{figure}
\centering
\includegraphics[width=9cm, angle=0]{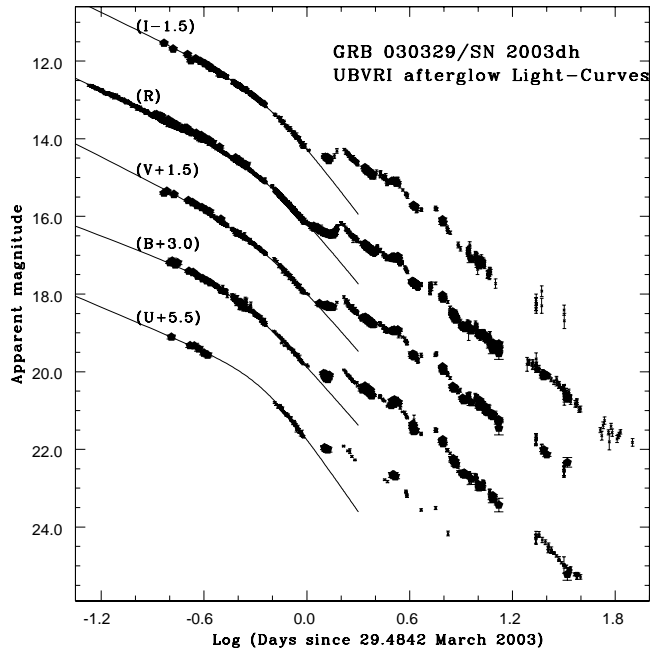}
\caption{Light-curves of GRB 030329/SN 2003dh showing the deviation from the 
modeled light-curve of afterglows (Pandey 2006). The light-curve at late epochs 
are flattened due to contribution of the underlying SN 2003dh along with
other light curve features. The time-resolved observations of such events
could tell a lot about the explosion mechanism and underlying physical
conditions.}  
%\label{fig:Sample}
\end{figure}
%%%%%%%%%%%%%%%%%%%%%%%%%%%%%%%%%%%%%%%%%%%%%%%%%%%%%%%

Some of the broad-lined Tpye Ib/c SNe are characterized by very high kinetic 
energies (Iwamoto et al. 1998; Nomoto et al 2004; Mazzali et al. 2008),
have been observed to be associated both temporally and specially with long 
duration GRBs. Apart from spectroscopically confirmed such associations, 
several late time red-bumps have also been identified in the late time 
afterglow light-curves of long duration GRBs, close resemblance with SN 1998bw 
profile (See Figure 3 above; Cano et al. 2011). The GRBs associated Type Ib/c SNe couple 
bulk of their 
energy to relativistic ejecta whereas ordinary SNe Ib/c couple less than a 
percent of their total energy to the fast material. Type Ib/c SNe related 
with GRBs have higher ejected mass in comparison to those not associated 
with GRBs (Mazzali et al. 2005,2006; Taubenberger et al. 2006; Valenti et al. 
2008). Nature of the progenitors of SNe of Type Ib/c is still not understood 
and that how do they differ from those associated with GRBs. 
However, the link to GRBs is a strong hint that GRBs associated 
Type Ib/c SNe 
could be significantly aspherical. If a jet is produced by a collapsing 
star (MacFadyen \& Woosely 1999), then only it can emerge and generate 
a GRB. A jet like explosion is required for GRBs from energetics 
considerations, indicating asphericity in the explosion mechanism. 
Ordinary Type Ib/c SNe are 
clearly distinguishable from the ones associated with the GRBs in terms 
of the relativistic ejecta produced by the central engine. There are some 
cases of low redshift Type Ib/c SNe with broad lines typical of 
hypernovae but not always associated with large luminosities typical of 
hypernovae (Mazalli et al. 2005; Valenti et al. 2008; Sahu et al. 2009; 
Drout et al. 2011). In the local universe, 
it has been seen that the rate of energetic broad-lined Ic SNe is similar to 
the rate of long-duration GRBs, which might indicate that most (or all) 
energetic Ic SNe produce GRBs (Podsiadlowski et al. 2004). However, the 
observed rate of production of WR stars (initial masses of $>$ 40 M$_{\odot}$) 
in galaxies are much higher than the rate of broad-lined SNe of Type Ic, 
indicate that not all WR stars produce broad-lined Type Ib/c SNe. 
Associations of Type Ib/c SNe have mostly been seen with low-luminosity 
nearby long duration GRBs (Hjorth \& Bloom 2011; Melandri et al. 2012; 
Sanders et al. 2012). Whether cosmological GRBs at 
higher redshifts are associated with SNe are quite challenging due to 
contamination by the afterglow and the host galaxy and is an open question 
to be addressed with TMT and other such facilities in near future.    

\subsection{Host galaxies of GRBs} 

GRBs being far away, lack direct identification of their progenitor stars. The 
evidence of what gives rise to GRBs therefore remains circumstantial. One 
important input to this body of evidence comes from the study of their host 
galaxies, and wherever possible, that of the immediate stellar neighborhood 
of the GRB. The GRB sample goes out to high redshift: the measured GRB 
redshifts at present have a median value of z $\sim$ 2.7. The host 
galaxies of GRBs 
that have been detected to date have their redshifts in the range 0.01--8.3. 
Clearly, a large fraction of the high redshift GRB host population remains 
to be discovered. The majority of the known hosts have magnitudes in the range 
R $\sim$ 22-27 mag and K $\sim$ 20-25 mag, close to the limit of studies possible with the 
present-day 8-10m class telescopes (Savaglio et al. 2009). The TMT and the 
proposed back-end instruments will have an excellent opportunity to address 
questions exploring  nature of GRB hosts. In addition to increasing the sample 
size at the high redshift end, TMT will enable detailed spectroscopic studies 
of the host galaxies. At present multi-band photometry of 
GRB hosts is being used to fit synthetic spectral models, from which the mass, 
metallicity, age and star formation rates are being deduced. The derived 
values suffer from very large uncertainties, making it difficult to reach 
definite conclusions about the nature of these galaxies and their population. 
TMT with the combination of OA and IFU will enable to study these objects 
spectroscopically, enhancing the reliability of model fitting and parameter 
extraction manifold. The present viewpoint that GRBs originate in sub-solar 
luminosity, actively star-forming galaxies with low metallicity can be 
seriously tested only by such detailed spectroscopic study about these host 
galaxies. Extending this study with TMT to a larger sample, especially at high 
redshifts would be able to tell more about the correlation of GRBs with 
low-metallicity progenitors. Morphological classification of GRB hosts, 
carried out mainly using the {\it HST} (Fruchter et al. 2006; Wainwright et al. 2007)
so far, seems to suggest a preponderance of irregulars among them. The 
enhanced angular resolution of the TMTs compared to the present day ground 
based telescopes can be put to use in conducting such morphological 
classification of a larger sample of GRB hosts and their mass function. 
Multi-band imaging with TMT could be used to understand the morphology of 
long and short duration host galaxies more in detail. This kind of study 
may also throw some light on whether the GRB production rate is influenced 
by galaxy mergers and the nature of galaxies hosting ``dark GRB''. 

\section{Polarimetric studies of  CCSNe and GRBs}

It is now established that afterglows of GRBs arise from synchrotron emission 
in a relativistically shocked ejecta, collimated outflow and 
emission is intrinsically polarized (Lazzati 2006; Wang \& Wheeler 2008). 
Evolution in the polarization as a function of viewing geometry as well 
as degree of orientation of the magnetic field (Medvedev \& Loab 1999) 
is expected in collimated outflows of GRBs. Models 
show that the maximum degree of polarization occurs around the time of 
geometric break called "jet-break" (Rossi et al. 2004). For GRB afterglows, 
optical imaging polarimetry have been successfully done in case of handful 
of GRBs and a polarization upto ~ 9\% have been observed 
(Bersier et al. 2003, Wiersema et al. 2012a). For the observed afterglows in 
optical-NIR, lack of strong variability in the position angle of linear 
polarization indicate that magnetic field of the jet to be ordered rather 
than a random one. Even with the help of existing 8-10m class telescopes 
like Keck and VLT, spectropolarimetric observations of GRB afterglows have 
been covered sparsely with no concluding results (Wiersema et al. 2012b). 
Some of the Type Ib/c SNe associated 
with GRBs have also shown polarization indicating towards asymmetry of the 
ejecta (photosphere/chemical inhomogeneity) for early photospheric phase of 
SNe (Kawabata et al. 2002, 2003; Tanaka et al. 2009). So far, spectropolarimetry of 
core-collapse SNe have only been possible using 8-10m class telescopes with 
rather low-resolutions (Hoffman et al. 2008) not providing any conclusive 
results about the structure of magnetic fields, rotational distortion of the 
progenitor stars and effect of interaction between circumstellar material 
and the ejected matter. 

Bigger observing facilities like TMT will be able to collect many more 
photons, enabling to measure the polarization more accurately. However, 
the present design of the TMT with a three-mirror system is expected to produce 
considerable instrumental polarization due to tertiary mirror at Nasmyth 
focus (Tinbergen 2007) at optical-IR frequencies. To reduce the instrumental 
polarization some ideas have been suggested, like, putting a polarization 
modulator or an additional optical component in the optical path to 
cancel out the instrumental polarization. 
However, it is not easy to implement these possible solutions as the 
instrument will be rotating with the telescopes. To compensate the rotation 
of the telescope polarization, the possible ways are to put a achromatic 
half-wave plate or to bend the optical path using Fresnel Romb before the 
focal plane. It is highly required to optimize the design of TMT or other 
upcoming extremely large optical-IR telescopes to reduce the instrumental 
polarization to make best use of the collected photons from various faint 
celestial objects. Early phase spectropolarimetry of afterglows of GRBs 
and CCSNe will be able to tell us more about the explosion geometry, 
chemical structure, velocity, and composition of the SN ejecta.  

\section{Conclusions}

The TMT and the back-end instruments are capable to address some of the un-answered
questions about CCSNe and GRBs and their implications to understand the 
evolution of massive stars and star formation at high redshifts. Time resolved
optical-IR spectroscopy of PISN and CCSNe of type II at z $>$ 2 will be able to 
understand the shock-break out phase and interaction of ejecta with the 
circum-stellar matter in more details. Photometric identification of CCSNe at 
z $\sim$ 2--6 will be helpful to know about evolution of massive stars at high-z.
Time resolved observations of of afterglows of GRBs would be quite helpful not only
in understanding progenitors at high-z and population of massive stars but also to prob 
the ISM and IGM in detail. Possible polarimetric observations using TMT would be 
able to tell about the geometry of the ejecta and structure of the magnetic field 
in these energetic cosmic explosions.

\section*{Acknowledgments}
The author is thankful to organizers of TMT Science and Instrumentation workshop, 10-12 December 2012, Pune, India, to invite me to deliver the talk on SNe and GRBs. The author acknowledge the TMT documentation at www.tmt.org/documents/ while preparing this manuscript. The author is also thankful to Profs. Ram Sagar, D. Bhattacharya and G. C. Anupama to have a discussion on SNe and GRBs.


\begin{thebibliography}{}
\bibitem{} Anderson, J. P., Covarrubias, R. A., James, P. A. et al., 2010, {\it MNRAS}, {\bf 407}, 2660
\bibitem{} Anderson, J. P., Habergham, S. M., James, P. A. et al., 2012, {\it MNRAS}, {\bf 424}, 1372
\bibitem{} Anderson, J.~P. \& James, P.~A., 2009, {\it MNRAS}, {\bf 399}, 559
\bibitem{} Arcavi, I., Gal-Yam, A., Kasliwal, M. M. et al., 2010, {\it ApJ}, {\bf 721}, 777
\bibitem{} Ashok, B. N., Anupama, G. C., Prabhu, T. P. et al., 1987, {\it JA\&A}, {\bf 8}, 195
\bibitem{} Barkat, Z., Rakavy, G., \& Sack, N., 1967, {\it Phys. Rev. Lett.}, {\bf 18}, 379
\bibitem{} Berger, E., Kulkarni, S. R., Bloom, J. S. et al., 2002, {\it ApJ}, {\bf 581}, 981
\bibitem{} Bersier, D., Stanek, K. Z., Winn, J. N., et al. 2003, {\it ApJ}, {\bf 584}, L43
\bibitem{} Bhattacharya, D., 2003, {\it BASI}, {\bf 29}, 107 
\bibitem{} Boissier, S. \& Prantzos, N., 2009, {\it A\&A}, {\bf 503}, 137 
\bibitem{} Bufano, F., Immler, S., Turatto, M., et al., 2009, {\it ApJ}, {\bf 700}, 1456
\bibitem{} Castro-Tirado, A. J., Zapatoras-Osorio, M. R., Caon, N. et al., 1999, {\it Science}, {\bf 283}, 2069
\bibitem{} Cano, Z., Bersier, D.; Guidorzi, C. et al., 2011, {\it ApJ}, {\bf 740}, 41
\bibitem{} Chatzopoulos, E. \& Wheeler, J. C., 2012, {\it ApJ}, {\bf 760}, 154
\bibitem{} Chevalier, R. A. \& Fransson, C., 2006, {\it ApJ}, {\bf 651}, 381
\bibitem{} Cooke, J., Coory, A., Chary, R. et al., 2009, {\it White paper Astro2010 decadle survey}, {\bf arXiv:0902:4602} 
\bibitem{} Crockett, R. M., Eldredge, J. J., Smartt, S. J., et al., 2008, {\it MNRAS}, {\bf 391}, L5
\bibitem{} Crockett, R. M., Smartt, S. J., Eldredge, J. J., et al., 2007, {\it MNRAS}, {\bf 381}, 835 
\bibitem{} Crompton, D., Simard, L. \&  Silva, D., 2008, {\it proceedings of the ESO Workshop ``Science with the VLT in the ELT Era''}, {\bf arXiv:0801:3634}
\bibitem{} Crowther,  P. A., 2007, {\it ARA\&A}, {\bf 45}, 177
\bibitem{} Crowther, P. A., 2013, {\it submitted to MNRAS}, {\bf arXiv:1210:1126}  
\bibitem{} D\'Elia, V., Campana, S., Covino, S., et al., 2011, {\it MNRAS}, {\it 418}, 680
\bibitem{} D\'Elia, V., Fiore, F., Meurs, E. J. A., 2007, {\it A\&A}, {\bf 467}, 629 
\bibitem{} D\'Elia, V., Fiore, F., Perna, R., et al., 2009, {\it A\&A}, {\bf 503}, 437
\bibitem{} Drout, M. R., Soderber, A. M., \& Gal-Yam, A. et al., 2011, {\it ApJ}, {\bf 741}, 97
\bibitem{} Eldridge, J. J., Fraser, M., Smartt, S. J., et al., 2013, {\it submitted to MNRAS}, {\bf arXiv:1301:1975}
\bibitem{} Fesen, R. A., Gerardy, C. L., Fillipenko, A. V. et al., 1999, {\it AJ}, {\bf 117}, 725
\bibitem{} Filippenko, A.~V., 1997, {\it ARA\&A}, {\bf 35}, 309
\bibitem{} Frail, D. A., Kulkarni, S. R., Sari, R., 2001, {\it ApJ}, {\bf 562}, L55
\bibitem{} Fryer, C.~L., 1999, {\it ApJ}, {\bf 522}, 413 
\bibitem{} Fryer, C. L. \& Heger, A., 2005, {\it ApJ}, {\bf 623}, 302 
\bibitem{} Fruchter, A. S., Levan, A. J., \& Strolger, L. et al. 2006, {\it Nature}, {\bf 441}, 463, 
\bibitem{} Fynbo, J. P. U., Jensen, B. L., Gorosabel, J. et al., 2001, {\it A\&A}, {\bf 369}, 373
\bibitem{} Ghirlanda, G., Ghisellini, G., Lazzati, G., et al., 2004, {\it ApJ}, {\bf 613}, L13
\bibitem{} Gehrels, N., Ramirez-Ruiz, E., Fox D.B., 2009, {\it ARA\&A}, {\bf 47}, 567
\bibitem{} Habergham, S. M., James, P. A. \& Anderson, J. P., 2012, {\it MNRAS}, {\bf 424}, 2841
\bibitem{} Heger, A., Fryer, C.~L. \& Woosley, S.~E., 2003, {\it ApJ}, {\bf 591}, 288 
\bibitem{} Heger, A \&  Woosley, S. E., 2002, {\it ApJ}, {\bf 567}, 532
\bibitem{} Hjorth, J. \& Bloom, J., 2011, {\it Chapter 9 in "Gamma-Ray Bursts", eds. C. Kouveliotou, R. A. M. J. Wijers, S. E. Woosley, Cambridge University Press, 2011}
Journal-ref: Gamma-Ray Bursts, {\bf Cambridge Astrophysics Series 51}, 169   
\bibitem{} Hoffman, J. L., Leonard, D. C., Chornock, R. et al., 2008, {\it ApJ}, {\bf 688}, 1186
\bibitem{} Iwamoto, K., Mazzali, P. A., Nomoto, K., 1998, {\it Nature}, {\bf 395}, 672
\bibitem{} Jakobasson, P., Hjorth, J., Fynbo, J. P. U. et al., 2004, {\it ApJ}, {\bf 617}, L21
\bibitem{} Joggerst, C. C. \& Whalen, D. J., 2011, {\it ApJ}, {\bf 728}, 129  
\bibitem{} Kasen, D., Wooley, S. E., Heger, A., 2011, {\it submitted to ApJ}, {\bf arXiv:1101.3336}
\bibitem{} Kawabata, K. S., Jeffery, D. J.; Iye, M. et al., 2002, {\it ApJ}, {\bf 580}, L39
\bibitem{} Kawabata, K. S., Deng, J., Wang, L. et al., 2003, {\it ApJ}, {\bf 593}, L19
\bibitem{} Kelly, P. K. \& Krishner, R. P., 2012, accepted to {\it ApJ}, {\bf arXiv:1110.1377} 
\bibitem{} Kumar, B., Pandey, S. B., Sahu, D. K., et al. 2013, {\it MNRAS}, {\bf 431}, 308 
\bibitem{} Lamb, D. Q. \& Reichart, D. E., 2000, {\it ApJ}, {\bf 536}, 1
\bibitem{} Langer, N., 2012, {\it ARA\&A}, {\bf 50}, 107 
\bibitem{} Langer, N. \& Norman, C. A., 2006, {\it ApJ}, {\bf 638}, L63 
\bibitem{} Langer, N., Norman, C. A., de Koter, A. et al., 2007, {\it A\&A}, {\bf 475}, L19
\bibitem{} Lazzati, D., 2006, {\it NJPh}, {\bf 8}, 131
\bibitem{} Lazzati, D., Covino, S., Alighieri, di S. et al., 2003, {\it A\&A}, {\bf 410}, 823 
\bibitem{} Leibundgut, B., Krishner, R. P., Pinto, P. A. et al., 1991, {\it ApJ}, {\bf 372}, 531 
\bibitem{} Lennarz, D., Altmann, D. \& Wiebusch, C., 2012, {\it A\&A} , {\bf 538}, 120
\bibitem{} MacFadyen, A. I. \& Woosley, S. E., 1999, {\it ApJ}, {\bf 524}, 262
\bibitem{} Maund, J. R., Smartt, S. J. \& Schweizer, F., 2005, {\it ApJ}, {\bf 630}, L33 
\bibitem{} Mazzali, P. A., Kawabata, K. S., Maeda, K. et al., 2005, {\it Science}, {\bf 308}, 1284 
\bibitem{} Mazzali, P. A., Deng, J., Pian, E. et al., 2006, {\it ApJ}, {\bf 645}, 1323 
\bibitem{} Mazzali, P. A., Valenti, S., Della Valle, M. et al. 2008, {\it Science}, {\bf 321}, 1185 
\bibitem{} Medvedev, M. V. \& Loeb, A., 1999, {\it ApJ}, {\bf 526}, 697
\bibitem{} Melandri, A., Pian E., Ferrero P., et al. 2012, {\it A\&A}, {\bf 547}, A82
\bibitem{} Malendri, A., Sbarufatti, B., D'Avanzo, P. et al., 2012, {\it MNRAS}, {\bf 421}, 1265
\bibitem{} M\'esz\'aros, P., 2002, {\it ARA\&A}, {\bf 40}, 137
\bibitem{} Milisavljevic, D., Fesen, R. A., Chevalier, R. A. et al., 2012, {\it ApJ}, {\bf 751}, 25
\bibitem{} Modjaz, M., Kewley, L., Bloom, J. S. et al., 2011, {\it ApJ}, {\bf 731}, L4 
\bibitem{} Modjaz, M. Kewley, L., Krishner, R. P. et al., 2008, {\it AJ}, {\bf 135}, 1136
\bibitem{} Murphy, J. W., Jennings, Z. G., Williams, B. el al., 2011, {\it ApJ}, {\bf 742}, 4 
\bibitem{} Neill, J. D., et al. 2011, {\it ApJ}, {\bf 727}, 15
\bibitem{} Nomoto, K. I., Iwamoto, K. \& Suzuki, T., 1995, {\it PhR}, {\bf 256}, 173
\bibitem{} Pandey, S. B., 2006, {\it PhD Thesis}
\bibitem{} Pandey, S. B., Anupama, G. C., Sagar, R. et al. 2003, {\it MNRAS}, {\bf 340}, 375 
\bibitem{} Pandey, S. B., Sahu, D. K., Resmi, L. et al., 2003, {\it BASI}, {\bf 31}, 19
\bibitem{} Pandey, S. B., Sagar, R., Anupama, G. C. et al., 2004, {\it A\&A}, {\bf 417}, 919
\bibitem{} Podsiadlowski, Ph, Joss, P. C., \& Hsu, J. J. L., 1992, {\it ApJ}, {\bf 391}, 246
\bibitem{} Podsiadlowski, Ph, Mazzali, P. A., \& Namoto, K. et al., 2004, {\it ApJ}, {\bf 607}, L17 
\bibitem{} Poznanski, D., 2013, {\it arXiv:1304:4967} 
\bibitem{} Poznanski, D., Nugent, P. E., Filippenko, A. V., 2008, {\it accepted to ApJ}, {\bf arXiv:1008:0877}
\bibitem{} Piran, T., 1999, {\it Physics Reports}, {\bf 314}, 575
\bibitem{} Piranomonte, S., Ward, P. A., Fiore, F., et al., 2008, {\it A\&A}, {\bf 492}, 775 
\bibitem{} Prochaska, J. X., Sheffer, Y., Perley, D. A. et al., 2009, {\it ApJ}, {\bf 691}, L27
\bibitem{} Quimby, R. M., Yuan, F., Akerlof, C. et al., 2013, accepted to MNRAS, {\it arXiv:1302:0911}
\bibitem{} Racusin, J. L., Karpov, S. V., Sokolowski, M. et al., 2008, {\it Nature}, {\bf 455}, 183 
\bibitem{} Rakavy, G. \& Shaviv, G., 1967, {\it ApJ}, {\bf 148}, 803
\bibitem{} Resmi, L., Misra, K., Jóhannesson, G. et al., 2013, {\it MNRAS}, {\bf 427}, 288
\bibitem{} Reichart, D. E. \& Price, P. A., 2002, {\it ApJ}, {\bf 565}, 174
\bibitem{} Rossi, E. M., Lazzati, D., Salmonson, J. D. et al., 2004, {\it ApJ}, {\bf 354}, 86
\bibitem{} Roy, R., Kumar, B., Benetti, S. et al., 2011, {\it ApJ}, {\bf 736}, 76
\bibitem{} Sagar, R., Kumar, B., Omar A., et al., 2012, {\it SPIE}, {\bf 8444}, 1TS
\bibitem{} Sagar, R., Pandey, A. K., Mohan, V., et al. 1999, {\it BASI}, {\bf 27}, 3  
\bibitem{} Sahu, D. K., Anupama, G. C., Srividya, S. et al. 2006, {\it MNRAS}, {\bf 372}, 1315 
\bibitem{} Sahu, D. K., Tanaka, M., Anupama, G. C. et al., 2009, {\it ApJ}, {\bf 697}, 676  
\bibitem{} Salvaterra, R., Della Valle, M., Campana, S., et al., 2009, {\it Nature}, {\bf 461}, 1258
\bibitem{} Sanders, N. E., Soderberg, A. M., Levesque, E. M., et al., 2012, {\it ApJ}, {\bf 758}, 132 
\bibitem{} Savaglio, S., 2006, {\it New Journal of Physics}, {\bf 8}, 9, 195 
\bibitem{} Savaglio, S., Glazebrook K., Le Borgne D., 2009, {\it ApJ}, {\bf 691}, 182
\bibitem{} Sari, R., Piran, T., Narayan, R., 1998, {\it ApJ}, {\bf 497}, L17 
\bibitem{} Scannapieco, E., Madau, P., Woosely, S. et al., 2005, {\it ApJ}, {\bf 633}, 1031
\bibitem{} Schawinski, K. Stephen, J., Christian, W. et al., 2008, {\it Science}, {\bf 321}, 223
\bibitem{} Smartt, S.~J., 2009, {\it ARA\&A}, {\bf 47}, 63
\bibitem{} Smartt, S. J., Eldridge, J. J., Crockett, R. M., et al., 2009,{\it  MNRAS}, {\bf 395}, 1409 
\bibitem{} Soderberg, A. M., Chakraborti, S., Pignata, G., 2010, {\it Nature}, {\bf 463}, 513 
\bibitem{} Soderberg, A. M., Kulkarni, S. R., Price, P. A. et al., 2006, {\it ApJ}, {\bf 636}, 391
\bibitem{} Steele, I. A., Mundell, C. G., Smith, R. J. et al., 2009, {\it Nature}, {\bf 462}, 767 
\bibitem{} Tanaka, M., Moriya, T. J., Yoshida H., et al.,  2012, {\it MNRAS}, {\bf arxiv:1202:3610}
\bibitem{} Tanvir, N. R., Fox, D. B., Levan, A. J., et al., 2009, {\it Nature}, {\bf 461}, 1254
\bibitem{} Taubenberger, S., Valenti, S., Benetti, S., et al., 2009, {\it MNRAS}, {\bf 397}, 677
\bibitem{} Taubenberger, S., Pastorello, A., Mazzali, P. A. et al., 2006, {\it MNRAS}, {\bf 371}, 1459 
\bibitem{} Tinbergen, J., 2007, {\it PASP}, {\bf 117}, 1371
\bibitem{} TMT\_documants; {\it TMT.PMO.MGT.07.009}, Editors: Dawson, S., Roberts, S.
\bibitem{} TMT\_documants; {\it TMT.PSC.TEC.07.003.REL01}, Editors: Silva, D., Hickson, P., Steidel, C., Bolte, M.
\bibitem{} TMT\_documants; {\it TMT.PSC.DRD.05.001.CCR18}, Editors: Nelson, J.
\bibitem{} Tominaga, N., Morokuma, T., Blinnikov, S. I., et al.,  2011, {\it ApJS}, {\bf 193}, 20
\bibitem{} Tout, C. A., Wickramasinghe, D.T., Lau, H.H-B. et al., 2011, {\it MNRAS}, {\bf 410}, 2458
\bibitem{} Valenti, S., Benetti, S., Cappellaro, E. et al., 2008, {\it MNRAS}, {\bf 383}, 1485
\bibitem{} van der Horst, A. J., et al. 2009, {\it ApJ}, {\bf 699}, 1087
\bibitem{} Vreeswijk, P. M., Ledoux, C., Raassen, A. J. J. et al., 2013, {\it A\&A}, {\bf 549}, 22	
\bibitem{} Vreeswijk, P. M., Ledoux, C., Smette, A. et al., 2007, {\it A\&A}, {\bf 468}, 83
\bibitem{} Ledoux, C., Vreeswijk, P. M., Smette, A. et al., 2009, {\it A\&A}, {\bf 506}, 661
\bibitem{} Wainwright, C., Berger, E., \& Penprase, B. E., 2007, {\it ApJ}, {\bf 657}, 367
\bibitem{} Wang, L. \& Wheeler, J. C., {\it ARA\&A}, {\bf 46}, 433 
\bibitem{} Whalen, D. J., Even, W., Lovekin, C. C. et al., 2013a, {\it submitted to ApJ}, {\bf arXiv:1302.0436}
\bibitem{} Whalen, D. J., Even, W., Frey, L. H. et al., 2013b, {\it submitted to ApJ}, {\it arXiv:1211.4979} 
\bibitem{} Whalen, D. J., Heger, H., Chen, K.~J. et al., 2013c, {\it submitted to ApJ}, {\it arXiv:1211.1815}
\bibitem{} Wiersema, K., Curran, P. A., Kruhler, T., et al. 2012, {\it MNRAS}, {\bf 426}, 2
\bibitem{} Wiersema, K., van der Horst, A. J., Levan, A. J. et al., 2012b, {\it MNRAS}, {\bf 421}, 1942
\bibitem{} Wijers, R. A. M. J. \& Galama, T. J., 1999, {\it ApJ}, {\bf 523}, 177 
\bibitem{} Woosley, S.E., Bloom, J.S., 2006, {\it ARA\&A}, {\bf 47}, 507
\bibitem{} Woosley, S. \& Janka, T., 2005, {\it NatPh}, {\bf 1}, 147
\bibitem{} Van dyk, S. D., Li, W., Cenko, S. B. et al., 2012, {\it ApJ}, {\bf 756}, 131
\bibitem{} Yoon, S. C., Langer, N., 2005, {\it A\&A}, {\bf 443}, 643
\bibitem{} Yoon, S. C., Grafener, G., Vink, J. S., et al., 2012, {\it A\&A}, {\bf 544}, L11
\bibitem{} Zhang, B., 2007, {\it ChJA\&A}, {\bf 7}, 1, 1 

\end{thebibliography}
\end{document}